\documentstyle[entcsmacro]{entcs}

\def\lastname{Dershowitz et al.}
\input{psfig}
\newcommand{\wgefig}[4]
        {\begin{figure}[htb] 
        \begin{center}%
        \mbox{}
        {\psfig{figure=#1,width=#4}}
        \mbox{}
        \end{center}
        \caption{{#2}\label{#3}} 
        \end{figure}}

\newcommand{\wrt}{w.r.t.}                       
\newcommand{\eat}[1]{}
\newcommand{\imp}{{\ :\!\!-\ }}
\newcommand{\mc}{{mc\_carthy\_91}}

\begin{document}

\begin{frontmatter}
\title{Automatic Termination Analysis of Programs Containing Arithmetic 
Predicates}
\author[Tel-Aviv]{Nachum Dershowitz}
\author[Jerusalem]{Naomi Lindenstrauss}
\author[Jerusalem]{Yehoshua Sagiv}
\author[Leuven]{Alexander Serebrenik}
\address[Tel-Aviv]{School of Mathematical Sciences\\ Tel-Aviv University\\ 
Tel-Aviv 69978, Israel}
\address[Jerusalem]{Institute for Computer Science\\The Hebrew University\\
Jerusalem 91904, Israel}
\address[Leuven]{Department of Computer Science\\Katholieke Universiteit Leuven\\
B-3001 Heverlee, Belgium}
\begin{abstract}
For logic programs with arithmetic predicates, showing termination 
is not easy, since the usual order for the integers is not well-founded. 
A new method, easily incorporated in the TermiLog system for automatic
termination analysis, is presented for showing termination in this case.

The method consists of the following steps:
First, a finite abstract domain for representing the range of integers
is deduced automatically. Based on this abstraction, abstract 
interpretation is applied to the program.  The result is a finite 
number of atoms 
abstracting answers to queries which are used to extend the 
technique of query-mapping pairs. For each query-mapping 
pair that is potentially non-terminating, 
a bounded (integer-valued) termination function is guessed.
If traversing the pair decreases the value of the termination function,
then termination is established.
Simple functions often suffice for each query-mapping
pair, and that gives our approach an edge over the classical approach of 
using
a single termination function for all loops, which must inevitably
be more complicated 
and harder to guess automatically.
It is worth noting that the termination of McCarthy's
91 function can be shown automatically using our method.

In summary, the proposed approach is based on combining 
a finite abstraction of the integers with the technique
of the query-mapping pairs, and is essentially capable of dividing 
a termination proof into several cases, such that a simple termination 
function suffices for each case. Consequently, the whole process of 
proving 
termination can be done automatically in the framework of TermiLog
and similar systems. 
\end{abstract}
\end{frontmatter}

\section{Introduction}
In studying termination of both pure logic programs and of real Prolog 
programs, we discovered that in most cases, 
termination of the programs we encountered
depended on the following factors:
\begin{enumerate}
\item {\em Simple structural recursion.} This case may usually be resolved by 
the use of  term size or list length as a norm~\cite{DeSchreye:Decorte:NeverEndingStory,Plumer:Book}. {\em General linear norms}, defined in~\cite{Lindenstrauss:Sagiv}, are a generalisation of these standard norms.
\item {\em Recursion with local variables.}
In this case an additional preprocessing step,
deriving interargument relations, is necessary~\cite{Benoy:King,Brodsky:Sagiv,Ullman:van:Gelder,Plumer}.
\item {\em Pseudo-recursion.} These are calls to recursively defined 
predicates 
that can be eliminated by repetitive unfolding~\cite{Apt:Book,Bossi:Cocco,Lindenstrauss:Sagiv:Serebrenik:L,Tamaki:Sato}.
\item {\em Non-logical features of Prolog.}
Such components of Prolog programs ({\tt assert/1, !/1, ->/2, findall/3}, etc.) 
have historically attracted 
less attention.  Termination \wrt\ control predicates may be found, for 
example, in~\cite{Lindenstrauss:Sagiv}. Termination \wrt\ cut was studied
in~\cite{Andrews:WST99}.
\item {\em Numerical loops.}  They are the topic of this paper.
\item {\em Non-linear loops.}  These are situations in which recursion is not covered
 by general linear norms, 
as defined in~\cite{Lindenstrauss:Sagiv}.   
\end{enumerate}

Termination of logic programs in the general case is undecidable (see
\cite{Apt:Handbook} for the formal proof). However, the simple semantics
of logic programs made the search for sufficient conditions for termination
a challenge for the research community.

Research on the first two topics of the list above
led to completely automated tools for 
verifying termination~\cite{Codish:Taboch,Lindenstrauss:Sagiv:Serebrenik},
based on the use of linear norms (cf.\ ~\cite{Bossi:Cocco:Fabris,DeSchreye:Decorte:NeverEndingStory,Plumer:ICLP,Vershaetse:DeSchreye:Deriving:Linear:Size:Relations}).
These systems are powerful enough to deal with a large fraction 
of the programs that have appeared in the literature~\cite{Apt:Book,Maria:Benchmarks,DeSchreye:Decorte:NeverEndingStory,Lindenstrauss:Sagiv}.
Moreover, most of the examples can be proved using term-size or,
less often, list-size. 
When other linear norms are 
necessary, the user is expected to provide them.
For any given program, these tools either provide a termination proof, or else
report that there may be cases of non-termination.  

Automatic linear norm inference was
studied in~\cite{Decorte:DeSchreye}. However, there are
examples, for which (it can be proved that) 
no general linear norm can demonstrate 
termination. These examples are covered by the last two items in the list 
above.

In this paper we concentrate on Case 5. Our approach is suitable for 
automatization and may be integrated in existing systems. In Case 6
more sophisticated orderings (like recursive path ordering~\cite{Dershowitz}) 
or non-numerical sizes should be used. 
These may be incorporated in the query-mapping pairs technique,
as described in Section~\ref{QM}. The difficulty here is in discovering
them automatically, though techniques can be borrowed from the term-rewriting 
literature.

The remainder of the paper is organized as follows: In 
Section~\ref{Motivation} a motivating example is given. 
In Sections~\ref{Preliminaries}--\ref{QM} 
the different components of the algorithm are explained and
in Section~\ref{Algorithm} the complete algorithm is formulated. 
In Section~\ref{Conclusion} some conclusions are presented.

\section{The 91 function}
\label{Motivation}
We start by illustrating the use of our algorithm for proving the termination
of the 91 function. This deliberately contrived function was invented by John
McCarthy for 
exploring properties of recursive programs, and is considered to be a good
test case for automatic verification systems (cf. \cite{Giesl,Knuth,Manna:McCarthy}). The treatment here is on the intuitive level.
Formal details will be given in subsequent sections.

Consider the clauses:

\begin{example}
\label{91:function}
\begin{eqnarray*}
&&\mathtt {\mc(X,Y) \imp X>100, Y\;\;is\;\; X-10.}\\
&&\mathtt {\mc(X,Y) \imp X\leq 100, Z\;\;is\;\; X+11,\mc(Z,Z1),} \\
&&\hspace{4.1cm}\mathtt {\mc(Z1,Y).}
\end{eqnarray*}

\noindent
and assume that a query of the form {\tt \mc($i$,$f$)} is given, 
that is, a query in which
the first argument is bound to an integer, and the second is free. This
program computes the same answers as the following one:

\begin{eqnarray*}
&&\mathtt {\mc(X,Y) \imp X>100, Y\;\;is\;\;X-10.}\\
&&\mathtt {\mc(X,91) \imp X\leq 100.}
\end{eqnarray*}
\noindent
with the same query. Note, however, that while the termination of the latter
program is obvious, since there is no recursion in it, the termination of
the first one is far from being trivial and a lot of effort was dedicated
to find termination proofs for it (\cite{Giesl,Knuth,Manna:McCarthy}).
\end{example}

Our algorithm starts off by {\bf discovering numerical arguments}. This step
is based on abstract interpretation, 
and as a result both arguments of {\tt \mc} are proven to be numerical. 
Moreover,
they are proven to be of integer type. The importance of knowledge of this 
kind and 
techniques for its discovery are discussed in Subsection~\ref{Discovering}.

The next step of the algorithm is the inference of the {\bf integer 
abstraction domain} which will help overcome difficulties caused by the fact 
that the
(positive and negative) integers with the usual (greater-than or less-than) order are not well-founded. Integer abstractions are
derived from arithmetic comparisons in the bodies of rules. However,
a simplistic approach may be insufficient and the more powerful techniques
presented in Section~\ref{IntDomain} are sometimes essential. In our case 
the domain $\{(-\infty,89], [90,100], [101,\infty)\}$ of intervals is deduced. For the sake of
convenience we denote this tripartite domain by $\{${\sl small, med, big}$\}$.

In the next step, we {\bf use abstract interpretation to describe answers to 
queries}. This allows us to infer numerical inter-argument relations of a 
novel type.  
In Section~\ref{AbstrInterp} the technique for inference of constraints of this
kind is presented. For our running example we get the following abstract atoms.

\medskip
\begin{tabular}{ll}
{\tt \mc({\sl big},{\sl big})}& {\tt \mc({\sl med},{\sl med})}\\
{\tt \mc({\sl big},{\sl med})}& {\tt \mc({\sl small},{\sl med})}
\end{tabular}
\medskip

\noindent These abstract atoms characterise the answers of the program.
 
The concluding step creates {\bf query-mapping pairs} in the fashion of~\cite{Lindenstrauss:Sagiv}. This process uses the abstract descriptions
of answers to queries and is described in Section~\ref{QM}.
In our case, we obtain among others, the query-mapping pair
having the query {\tt \mc($i$,$f$)} and the mapping presented in 
Figure~\ref{qm91a}. The upper nodes correspond to argument positions of the
head of the recursive clause, 
and lower nodes---to argument positions of the second
recursive subgoal in the body. 
Black nodes denote integer argument positions,
and white nodes denote positions
on which no assumption is made. The 
arrow denotes an increase of the first argument, in the sense that
the first argument in the head is less than 
the first argument in the second recursive subgoal. 
Each set of nodes is accompanied by a set of constraints. Some
are inter-argument relations of the type considered
in~\cite{Lindenstrauss:Sagiv}. In our example this subset is empty. The rest
are constraints based on the integer abstraction domain. In this
case, that set contains the constraint that the first argument is in
{\sl med}. 
\wgefig{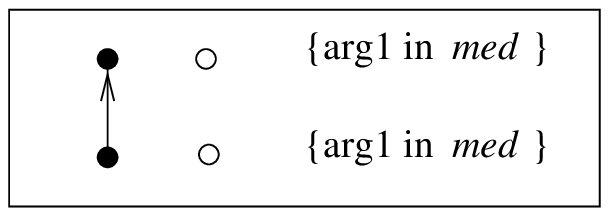}{Mapping for McCarthy's 91 function}{qm91a}{0.40\textwidth}
The query-mapping pair presented is circular (upper and lower nodes are the same
),
but the termination tests of~\cite{Codish:Taboch,Lindenstrauss:Sagiv} fail.  
Thus, a termination function must be guessed.
For this loop we can use the function $100-${\sl arg1}, where 
{\sl arg1} denotes the first argument of the atom.
The value of this function decreases while traversing the given 
query-mapping pair from the upper to the lower nodes.
Since it is also bounded from below ($100\geq ${\sl arg1}),
this query-mapping pair may be traversed only finitely many times. The
same holds for the other circular query-mapping pair in this case. 
Thus, termination is proved.

\section{Logic programs containing arithmetic predicates}
\label{Preliminaries}

The algorithm we describe here would come into play
only when the usual termination analysers fail to prove
termination using the structural arguments of predicates. 
As a first step it verifies the presence of an integer loop in 
the program. If no integer  loop is found, the possibility of 
non-termination is reported, meaning that the termination
cannot be proved by this technique. If integer loops are found, each of them 
is taken into consideration. 
The algorithm starts by discovering integer positions in the program, 
proceeds with creating appropriate abstractions, based on the integer loops,
and concludes by applying an extension of the query-mapping pairs technique. 
The formal algorithm is presented in Section~\ref{Algorithm}.

\subsection{Numerical and integer loops}
Our notion of numerical loop is based on the predicate dependency graph
(cf.\ \cite{Plumer:Book}):

\begin{definition}
Let $P$ be a program and let $\Pi$ be a strongly connected component
in its predicate dependency graph.  Let $S\subseteq P$ be the set of 
program clauses, associated with $\Pi$ (i.e. those clauses that have the 
predicates of $\Pi$ in their head).
$S$ is called {\em loop} if there is a cycle through predicates of $\Pi$.
\end{definition}
\begin{definition}
A loop $S$ is called {\em numerical\/} if there 
exists $H\imp B_1,\ldots,B_n$ in $S$, such that for some $i$, 
$B_i\equiv \mbox{\sl Var\ }\mbox{\tt is\ }\mbox{\sl Exp}$,
and either {\sl Var} is equal to some argument of $H$ or {\sl Exp} is an arithmetic expression involving a variable that is equal to some argument of $H$.
\end{definition}

However,
termination of numerical loops that involve numbers that are not integers
often depends on the specifics of
implementation and hardware, so
 we  limit ourselves
to ``integer loops'', rather than all numerical loops. 
The following examples illustrate actual behaviour that
contradicts intuition of general numerical loops---a loop
that should not terminate terminates, while a loop that 
should terminate does not. We checked the
behaviour of these examples on UNIX, with 
the CLP(Q,R) library~\cite{CLP:Manual} of SICStus Prolog~\cite{SICStus:Manual},
CLP(R)~\cite{Jaffar:Maher} and XSB~\cite{XSB:Manual}.

\begin{example}
\label{ex:real:loops}
Consider the following program. The goal {\tt p(1.0)} terminates although we
would expect it not to terminate. On the other hand the goal {\tt  q(1.0)}
does not terminate, although we would expect it to terminate.
\begin{eqnarray*}
&& \mathtt {p(0.0)\imp !.}\\
&& \mathtt {p(X) \imp X1\;\;is\;\;X/2,\;\;p(X1).}\\
&&\\
&& \mathtt {q(0.0)\imp !.}\\
&& \mathtt {q(X) \imp X1\;\;is\;\;X\;-\;0.1,\;\;q(X1).}
\end{eqnarray*}
\end{example}

One may suggest that assuming that the program does not contain division
and non-integer
constants will solve the problem. The following example shows that this
is not the case:
\begin{example}
\label{ex:num:loop}
\begin{eqnarray*}
&& \mathtt {r(0).}\\
&& \mathtt {r(X) \imp X>0,\;\;X1\;\;is\;\;X-1,\;\;r(X1).}
\end{eqnarray*}
The predicate {\tt r} may be called with a
real, non-integer argument, and then its behaviour is implementation dependent.
For example, one would expect that 
{\tt r(0.000000001)} will fail and  {\tt r(0.0)} will succeed. 
However, in SICStus Prolog both goals
fail, while in CLP(R) both of them succeed!
\end{example}

Therefore, we limit ourselves to integer loops, that is numerical loops 
involving integer constants and arithmetical calculations over integers:
\begin{definition}
A program $P$ is {\em integer-based} if, given a query such that all 
numbers appearing in it are integers, all subqueries that arise have this 
property as well.
\end{definition}

Although this definition may seem overly restrictive we use it to avoid 
unnecessary complications.

\begin{definition}
A numerical loop $S$ in a program $P$ is called an {\em integer\/} loop if $P$
is integer-based.
\end{definition}

Termination of a query may depend on whether its argument is an integer, as
the following example shows:

\begin{eqnarray*}
&& \mathtt {p(0).}\\
&& \mathtt {p(N) \imp N\;>\;0,\;\;N1\;\;is\;\;N\;-\;1,\;\;p(N1).}\\
&& \mathtt {p(a) \imp p(a).}
\end{eqnarray*}

\noindent For this program, {\tt p($n$)} for integer $n$ terminates, while 
{\tt p(a)} does not. 

So we extend our notion of query pattern.
Till now a  query pattern was an atom with the
same predicate and arity as the query, and arguments  $b$ 
(denoting an argument that is instantiated enough with respect to the norm) 
or $f$ (denoting an argument on which no 
assumptions are made). Here, we extend the notion to include 
arguments of the form $i$,
denoting an argument that is an integer (or integer expression). Note that $b$
includes the possibility of $i$, in the same way that $f$ includes the 
possibility $b$. In the diagrams to follow we denote $i$-arguments by 
black nodes, $b$-arguments by gray nodes and $f$-arguments by white nodes.

Our termination analysis is always performed with respect to a given program 
and a query pattern. A positive response guarantees termination of every query
that matches the pattern.  

\subsection{Discovering numerical arguments}
\label{Discovering}
Our analysis that will be discussed in the subsequent sections is based
on the size relationships between ``numerical arguments''. These are arguments
that are numerical for all subqueries generated from the initial query. 

The inference is done in two phases---bottom-up and top-down. 
Bottom-up inference is similar to type analysis 
(cf.\ \cite{Boye:Maluszynski,Codish:Lagoon}), 
using the abstract domain
$\{\mbox{\sl int},\mbox{\sl not\_int}\}$ and the observation that an argument 
may become {\sl int} only
if it is obtained from {\tt is/2} or is bound to an integer expression of 
arguments 
already found to be {\sl int} (i.e. the abstraction of $\mbox{\sl int}+\mbox{\sl int}$
is {\sl int}). 
Top-down inference is query driven and is similar to the ``blackening'' 
process, 
described in~\cite{Lindenstrauss:Sagiv}, 
only in this case the information propagated is being an integer expression 
instead of ``instantiated enough''. 

The efficiency of discovering numerical arguments may be improved
by a preliminary step of guessing the numerical argument positions. The
guessing is based on the knowledge that variables appearing in comparisons
or {\tt is/2}-atoms should be numerical. Instead of considering the whole 
program it is sufficient in this case to consider only clauses of
predicates having clauses with the guessed arguments
and clauses of predicates on which they depend. The 
guessing as a preliminary step becomes crucial when considering
``real-world'' programs that are large, while their numerical part is 
usually small.

\section{Integer Abstraction Domain}
\label{IntDomain}
In this subsection we present a technique that allows us to overcome the 
difficulties caused by the fact that
the integers with the usual order are not 
well-founded. Given a program $P$ we introduce a finite abstraction domain,
representing integers. The integer abstractions are derived from the subgoals
involving integer arithmetic positions. 

Let $S$ be a set of clauses in $P$, consisting of an integer loop and 
all the clauses for predicates on which the predicates of the integer loop 
depend. As a first step for defining the abstract domain we
obtain the set of comparisons ${\cal C}$ for the clauses in $S$.

More formally, we consider as a 
{\em comparison}, an atom of the form $t_1 \rho t_2$, such that 
$t_1$ and $t_2$ are either variables or constants and 
$\rho\in \{<,\leq,\geq,>\}$. Observe that we restrict ourselves
only to these atoms in order to ensure the
finiteness of ${\cal C}$. Note that by excluding $\not =$ and $=$ we
do not limit the generality of the analysis. Indeed if $t_1 \not = t_2$
appears in a clause it may be replaced by two clauses containing  
$t_1 > t_2$ and $t_1 < t_2$ instead of $t_1 \not = t_2$, 
respectively. Similarly, if the clause contains a subgoal $t_1 = t_2$, 
the subgoal may be replaced by two subgoals $t_1 \geq t_2, t_1 \leq t_2$.
Thus, the equalities we use in the examples to follow should be seen as 
a brief notation as above.

In the following subsections we present a number of techniques 
to infer ${\cal C}$ from the clauses of $S$.

We define ${\cal D}_p$ as the set of pairs $(p,c)$, for all satisfiable $c \in 2
^{{\cal C}_p}$. Here we interpret $c \in 2^{{\cal C}_p}$ as a conjunction
 of the comparisons in $c$ and the negations of the comparisons in
${\cal C}_p \setminus c$.
The abstraction domain ${\cal D}$ is taken as the union of the sets
${\cal D}_p$ for the recursive predicates $p$ in $S$.
Simplifying the domain may improve the running time of the analysis, however
it may make it less precise. 

\subsection{The simple case---collecting comparisons}

The simplest way to obtain ${\cal C}$ from the clauses of $S$ is
to consider the comparisons appearing in the bodies of recursive clauses
and restricting integer positions in their heads.

We would like to view ${\cal C}$ as a set of comparisons of head argument
positions. Therefore we assume in the simple case that $S$ is
{\em partially normalised}, that is, all head {\em integer\/} 
argument positions in clauses of $S$ are occupied by distinct variables. 
Observe that the assumption holds for all the examples considered so far.
This assumption will not be necessary with the more powerful technique 
presented in the next subsection. 

Consider
\begin{example}
\label{comparisons}
\begin{eqnarray*}
&&\mathtt {t(X)\imp X>5, X<8, X<2, X1\;is\;X+1, X1<5, t(X1).}
\end{eqnarray*}
Let {\tt t($i$)} be a query pattern for the program above. 
In this case, the first argument of {\tt t} is an integer argument.
Since {\tt X1} does not appear in the head of the first clause
{\tt X1<5} is not considered and,
thus, ${\cal C}= \{X>5,X<8,X<2\}$. We have in this example
only one predicate and the union is over the singleton set.
So, ${\cal D} = \{X<2,2\leq X\leq 5,5<X<8,X\geq 8\}$.
\end{example}

The following example evaluates the {\sl mod\/} function. 
\begin{example}
\label{mod}
\begin{eqnarray*}
&& \mathtt {mod(A,B,C) \imp A\geq B, B>0, D\;is\;A-B, mod(D,B,C).}\\
&& \mathtt {mod(A,B,C) \imp A< B, A \geq 0, A = C.}
\end{eqnarray*}
Here we ignore the second clause since it is not recursive. Thus, by 
collecting 
comparisons from the first clause, ${\cal C} _{\mbox{\tt mod}}= \{
arg1\geq arg2, arg2>0\}$ and thus,
by taking all the conjunctions of comparisons of ${\cal C}$
and their negations, we obtain ${\cal D}_{\mbox{\tt mod}} = 
\{(\mbox{\tt mod},\mbox{\sl arg1}\geq \mbox{\sl arg2} \;\&\; \mbox{\sl arg2}>0),
(\mbox{\tt mod},\mbox{\sl arg1}\geq \mbox{\sl arg2} \;\&\; \mbox{\sl arg2}\leq 0
),
(\mbox{\tt mod},\mbox{\sl arg1}< \mbox{\sl arg2} \;\&\; \mbox{\sl arg2}>0),
(\mbox{\tt mod},\mbox{\sl arg1}< \mbox{\sl arg2} \;\&\; \mbox{\sl arg2}\leq 0)\}
$.
\end{example}

However, sometimes the abstract domain obtained in this way is insufficient
for proving termination, and thus, should be refined. The domain may be
refined by enriching the underlying set of comparisons. Possible
ways to do this are using inference of comparisons instead of collecting them, 
or performing an unfolding, and applying the collecting or inference 
techniques to the unfolded program. 

\subsection{Inference of Comparisons}

As mentioned above, sometimes the abstraction domain obtained from 
comparisons appearing in $S$ is insufficient. Thus, we would like to refine it.
Instead of collecting comparisons, appearing in bodies of clauses,
we collect certain comparisons that are {\em implied} by bodies of clauses.
For example, {\tt X is Y+Z} implies the constraint
{\sl X=Y+Z} and {\tt functor(Term,Name,Arity)} implies {\sl Arity}$\geq 0$. 

As before, we restrict ourselves to recursive clauses (or clauses
recursive predicates depend on) and comparisons that
constrain integer argument positions of heads. 
Since a comparison that is contained in the body is implied by it, 
we always get
a superset of the comparisons obtained by the collecting technique, presented 
previously. The set of comparisons inferred depends on the power of the 
inference engine used (e.g. CLP-techniques may be used for this purpose). 

We define the abstract domain ${\cal D}$ as above. Thus, the 
granularity of the abstract domain also depends on the power of the inference
engine.

\subsection{Unfolding}

Unfolding (cf.\ \cite{Apt:Book,Bossi:Cocco,Lindenstrauss:Sagiv:Serebrenik:L,Tamaki:Sato}) allows us to generate
a sequence of abstract domains, such that each refines the previous.

More formally, let $P$ be a program and let 
$H\imp B_1,\ldots,B_n$ be a recursive rule in $P$. 
Let $P_1$ be the result of unfolding an atom $B_i$ in 
$H\imp B_1,\ldots,B_n$ in $P$. Let $S_1$ be a set of clauses in $P_1$, 
consisting of an integer loop and the clauses 
of the predicates on which the integer loop predicates depend.
More formally, if $S$ is an integer loop, then by using the standard notation 
of Apt~\cite{Apt:Book} we define $S_1$ to be 
$S\cup \{H\leftarrow B| (H\leftarrow B)\in P \wedge
\exists (H_1\leftarrow B_1)\in S,\; \mbox{\sl s.t. }
\mbox{\sl rel($H_1$)}\sqsupseteq \mbox{\sl rel($H$)}\}$.

Obtain ${\cal D}$ for the clauses of $S_1$ 
either by collecting comparisons from
rule bodies or by inferring them, and use it as a new abstraction domain
for the original program.
If the algorithm still fails to prove termination, the process of
unfolding can be repeated. Note, that for the cases
encountered in practice at most one step of unfolding is necessary. 

\begin{example}
Unfolding {\tt \mc(Z1,Z2)} in the recursive clause we obtain a new program
for the query {\tt \mc($i$,$f$)}
\begin{eqnarray*}
&& \mathtt {\mc(X,Y) \imp X>100,\;Y\;\;is\;\;X-10.}\\
&& \mathtt {\mc(X,Y) \imp X\leq 100,\;Z1\;\;is\;\;X+11, Z1>100,}\\
&& \mathtt {\hspace{1.6in} \;Z2\;\;is\;\;Z1-10,\mc(Z2,Y).}\\
&& \mathtt {\mc(X,Y) \imp X\leq 100,\;Z1\;\;is\;\;X+11, Z1\leq 100,}\\
&& \mathtt {\hspace{1.6in}Z3\;\;is\;\;Z1+11,\mc(Z3,Z4),}\\
&& \mathtt {\hspace{1.6in}\mc(Z4,Z2),\mc(Z2,Y).} 
\end{eqnarray*}
Now if we use an inference engine that is able to discover that {\tt X is Y+Z}
implies the constraint {\sl X=Y+Z}, we obtain the
following constraints on the bound head integer variable $X$
(for convenience we omit redundant ones):
From the first clause we obtain $X>100$. From the second clause---%
$X\leq 100$, and since $X+11>100$ we get $X>89$. Similarly, from the third
clause---$X\leq 89$. Thus, ${\cal C}=\{X\leq 89,X>89 \wedge X\leq 100, X>100\}$
Substituting this in the definition of ${\cal D}$, and removing inconsistencies
and redundancies, we obtain ${\cal D}=\{X\leq 89,X>89 \wedge X\leq 100, X>100\}$.
\end{example}

\subsection{Propagating domains}
\label{propagating}
The comparisons we obtain by the techniques presented 
above may restrict only {\sl some\/} subset of integer argument
positions. However, for the termination proof, information on integer arguments
outside of this subset may be needed. 
For example, as we will see shortly,
in order to analyse correctly {\tt \mc} we need to determine the domain for
the second argument, while the comparisons we have constrain
only the first one. Thus, we need some technique of {\sl propagating\/}
abstraction domains that we obtained for one subset of integer argument
positions to another subset of integer argument positions. Clearly,
this technique may be seen as a heuristic and it is inapplicable if there
is no interaction between argument positions.

To capture this interaction we draw a graph for each recursive numerical 
predicate, that has the numerical argument positions as vertices and edges 
between vertices that can influence each other. In the case of {\tt \mc} we 
get the graph having an edge between the first argument position and the 
second one.

Let $\pi$ be a permutation of the vertices of a connected component of this 
graph. Define $\pi {\cal D}$ to be the result of replacing each occurrence
of $arg_i$ in ${\cal D}$ by $arg_{\pi (i)}$. 
Consider the Cartesian product of all abstract domains 
$\pi {\cal D}$ thus obtained, discarding unsatisfiable conjunctions.
We will call this Cartesian product the {\em extended domain\/} and denote
it by ${\cal E} {\cal D}$. 
In the case of {\tt \mc} we get as 
${\cal E}{\cal D}$ the set of elements 
{\tt \mc($A$,$B$)}, such that $A$ and $B$ are in
 $\{\mbox{\sl small},\mbox{\sl med},\mbox{\sl big}\}$. 

More generally, when there are arithmetic relations (e.g.\ $Y = X+1$) 
between argument positions  ${\cal E}{\cal D}$ can contain
new subdomains that can be inferred from those in ${\cal D}$.

\section{Abstract interpretation}
\label{AbstrInterp}
In this section we use the integer abstractions obtained earlier to
classify, in a finite fashion, all possible answers to queries. This analysis
can be skipped in  simple cases (just as in TermiLog 
constraint inference can be skipped when not needed), but is necessary 
in more complicated cases, like {\tt \mc}. Most examples encountered in
practice do not need this analysis.

The basic idea is as follows: define an abstraction domain and perform a
bottom-up constraints inference. 

The abstraction domain that should be defined is a refinement of the 
abstraction domain we defined in Subsection~\ref{IntDomain}. 
There we considered only recursive clauses, since non-recursive clauses do not
affect the query-mapping pairs.
On the other hand, when trying to infer constraints that hold for 
answers of the
program we should consider non-recursive clauses as well. In this way
using one of the techniques presented in the previous subsection both 
for the recursive and the non-recursive clauses 
an abstraction domain $\tilde{\cal D}$ is obtained.
Clearly, $\tilde{\cal D}$ is a refinement of ${\cal D}$.  

\begin{example}
For {\tt \mc} we obtain that the elements of $\tilde{\cal D}$ are the intersections of the elements in
$ {\cal E}{\cal D}$ (see the end of Subsection~\ref{propagating})with the constraint in the non-recursive clause and its negation.
\end{example}

\begin{example}
\label{Dtilde}
Continuing the {\tt mod}-example we considered in Example~\ref{mod} and 
considering the non-recursive clause for {\tt mod} as well, we
obtain by collecting
comparisons that $\tilde{\cal C} = \{arg1\geq arg2, arg2>0, arg1< arg2,
arg3<arg2,arg1 \geq 0,
 arg1\leq arg3, arg1\geq arg3\}$ and, thus, $\tilde{\cal D}$ consists of all 
pairs $(\mbox{\tt mod},c)$ for $c$ a satisfiable
element of $2^{\tilde{\cal C}}$.
\end{example}

Given a program $P$, let $\cal B$ be the corresponding extended Herbrand base,
where we assume that arguments in numerical positions are integers. 
Let $T_{P}$ be the immediate consequence operator. 
Consider the Galois connection provided by the
abstraction function $\alpha: {\cal B} \rightarrow \tilde{\cal D}$ 
and the  concretization function $\gamma: \tilde{\cal D} \rightarrow {\cal B}$ 
defined as follows:
The abstraction $\alpha$ of an element in ${\cal B}$ is the pair from
 $\tilde{\cal D}$
that characterises it. The concretization $\gamma$ of an element in 
$\tilde{\cal D}$ is the set of all atoms in ${\cal B}$ that satisfy it.
Note that $\alpha$ and $\gamma$ form a Galois connection due to the 
disjointness of the elements of $\tilde{\cal D}$.

Using the Fixpoint Abstraction Theorem (cf.\ \cite{Cousot:Cousot}) we get that
\[\alpha \left ( \bigcup _{n=1}^{\infty}T_{P}^{i}(\emptyset ) \right ) \subseteq
 \bigcup _{n=1}^{\infty} (\alpha \circ T_{P}\circ \gamma )^{i}(\emptyset)\]
We will take a map $\mbox{\tt w}: \tilde{\cal D} \rightarrow\tilde{\cal D}$, that is
a {\em widening\/}~\cite{Cousot:Cousot} of 
$\alpha \circ T_{P}\circ \gamma$ and compute its
fixpoint. Because of the finiteness of $\tilde{\cal D}$ this fixpoint may be
computed bottom-up. 

The abstraction domain 
$\tilde{\cal D}$ describes all possible atoms in the extended 
Herbrand base ${\cal B}$. 
However, it is sufficient for our analysis to describe only
computed answers of the program, i.e., a subset of ${\cal B}$.
Thus, in practice, the computation of the fixpoint can sometimes be simplified
as follows:
We start with the constraints of the non-recursive clauses. Then we repeatedly
apply the recursive clauses to the set of the constraints obtained thus far, 
but abstract the conclusions to elements of ${\cal D}$. In this way we obtain 
a CLP program that is an abstraction of the original one. This holds in the next example.
The abstraction corresponding to the predicate {\tt p} is denoted
${\mathtt {p_w}}$.

\begin{example}
Consider once more {\tt \mc}. As claimed above we start from the non-recursive
clause, and obtain that
\begin{eqnarray*}
&& \mathtt {\mc_w(A, B) \imp \{A>100,B=A-10\}.}
\end{eqnarray*}
By substituting in the recursive clause of {\tt \mc} we obtain the following
\begin{eqnarray*}
&& \mathtt {\mc(X,Y) \imp X\leq 100,\;Z1\;\;is\;\;X+11, Z1>100,}\\
&& \mathtt {\hspace{1.5in} \;Z2\;\;is\;\;Z1-10, Z2>100, \;Y\;\;is\;\;Z2-10.}
\end{eqnarray*}
By simple computation we discover that in this case {\tt X} is 100,
and {\tt Y} is 91. However, in order to guarantee the termination of
the inference process we do not infer the precise constraint
$\{X = 100, Y = 91\}$, but its abstraction, i.e., an atom 
$\mathtt {\mc_w({med}, {med})}$. Repeatedly applying the
procedure described, we obtain an additional answer 
$\mathtt {\mc_w({small}, {med})}$. 

More formally, the following SICStus Prolog CLP(R) program performs the 
bottom-up construction of the abstracted program, as described above.
We use the auxiliary predicate {\tt in/2} to denote a membership in ${\cal D}$
and the auxiliary predicate {\tt e\_in/2} to denote a membership in 
the extended domain ${\cal E}{\cal D}$.
\begin{eqnarray*}
&& \mathtt {\imp use\_module(library(clpr)).}\\
&& \mathtt {\imp use\_module(library(terms)).}\\
&& \mathtt {\imp dynamic(\mc_w/2).}\\
&& \\
&& \mathtt {in(X,big) \imp \{X> 100\}.}\\
&& \mathtt {in(X,med) \imp \{X>89,X\leq 100\}.}\\
&& \mathtt {in(X,small) \imp \{X\leq 89\}.}\\
&& \\
&& \mathtt {e\_in((X,Y),(XX,YY))\imp in(X,XX), in(Y,YY).}\\
&& \\
&& \mathtt {\mc_w(X,Y) \imp \{X>100, Y=X-10\}.}\\
&& \\
&& \mathtt {assert\_if\_new((H\imp B)) \imp \backslash\!+ (clause(H1,B1),}\\
&& \mathtt {\hspace{6.0cm}unify\_with\_occurs\_check((H,B),(H1,B1))),}\\
&& \mathtt {\hspace{5.2cm}assert((H\imp B)). }\\
&& \\
&& \mathtt {deduce \imp \{X\leq 100,Z=X+11\},\mc_w(Z,Z1),}\\
&& \mathtt {\hspace{2.0cm} \mc_w(Z1,Y),e\_in((X,Y),(XX,YY)),}\\
&& \mathtt {\hspace{2.0cm} assert\_if\_new((\mc_w(A,B) \imp e\_in((A,B),(XX,YY)))),}\\
&& \mathtt {\hspace{2.0cm}deduce.}\\
&& \mathtt {deduce.}
\end{eqnarray*}
The resulting abstracted program is
\begin{eqnarray*}
&& \mathtt {\mc_w(A, B) \imp \{A>100,B=A-10\}.}\\
&& \mathtt {\mc_w(A, B) \imp e\_in((A,B), (med, med)).}\\
&& \mathtt {\mc_w(A, B) \imp e\_in((A,B), (small, med)).}
\end{eqnarray*}
Since we assumed that the query was of the form {\tt \mc($i$,$f$)} we can view
these abstractions as implications of constraints like
$\mbox{\sl arg1}\leq 89$  implies $89<\mbox{\sl arg2}\leq 100$. We also point 
out that the resulting abstracted program coincides with the results obtained
by the theoretic reasoning above.
\end{example}

As an additional example consider 
the computation of the {\sl gcd} according to Euclid's 
algorithm. Proving termination is not trivial, even if the successor notation 
is 
used. In~\cite{Lindenstrauss:Sagiv:Serebrenik:L} only applying a special 
technique allowed to do this.

\begin{example}
Consider the following program and the query {\tt gcd($i$,$i$,$f$)}.
\begin{eqnarray*}
&& \mathtt {gcd(X,0,X) \imp X>0.}\\
&& \mathtt {gcd(X,Y,Z) \imp  Y>0,  mod(X,Y,U),  gcd(Y,U,Z).}\\
&& \\                          
&& \mathtt {mod(A,B,C) \imp A\geq B, B>0, D\;is\;A-B, mod(D,B,C).}\\
&& \mathtt {mod(A,B,C) \imp A< B, A \geq 0, A = C.}
\end{eqnarray*}
In this example we have two nested integer loops represented by the predicates 
{\tt mod} and {\tt gcd}. We would like to use the information obtained from 
the abstract interpretation of {\tt mod}
to find the relation between the {\tt gcd}-atoms in the recursive clause.
Thus, during the bottom-up inference 
process we  abstract the conclusions to  elements of
${\tilde{\cal D}_{\mathtt mod}}$, as it was evaluated in Example~\ref{Dtilde}.
Using this technique  
we get that if $mod(X,Y,Z)$ holds then always $Z<Y$ holds, and this is what is needed to prove the
termination of $gcd$.
\end{example}

The approach presented in this subsection is
a novel approach compared to the inter-argument relations that were 
used previously in termination analysis~\cite{Benoy:King,Brodsky:Sagiv:2,Codish:Taboch,DeSchreye:Decorte:NeverEndingStory,Lindenstrauss:Sagiv,Mesnard:Ganascia,Plumer:Book,Ullman:van:Gelder,Vershaetse:DeSchreye,Vershaetse:DeSchreye:Deriving:Linear:Size:Relations}---instead of comparing 
the sizes of arguments the ``if \ldots then \ldots'' expressions are 
considered, making the analysis more powerful.

\section{Query-mapping pairs}
\label{QM}
We use the query-mapping pairs technique, presented
in~\cite{Lindenstrauss:Sagiv}, for reasoning about termination. 
The basic idea is to partition query evaluation
into ``simple steps'', corresponding to one rule application and then
compose them. The ``steps'' are
performed over a finite abstract domain. Circular ``steps'' are suspicious. 
If, for each  such circular step, we succeed  in showing a decrease of
some
argument position according to some integer linear norm, termination is proved, 
due to the
well-foundedness of the norm. Since the abstract domain is finite,
we have to check only a finite number of 
objects. 
Here, we extend the technique to programs having numerical arguments.
We also assume that an integer linear norm is defined for all arguments.

As presented in~\cite{Lindenstrauss:Sagiv}, 
queries are given as constrained abstract atoms. 
More formally, let  $\leftarrow A_1,\ldots, A_n \leadsto \ldots \leadsto\; \leftarrow B_1,\ldots, B_k$ be a partial branch in the LD-tree~\cite{Apt:Book}, 
and let $\theta$ be the composition of the substitutions along 
this branch. Assume also that the atom $B_1$ came into being from the
resolution on $A_1$. In the query-mapping pair corresponding to this
branch, the query is the abstraction of $A_1$ and the
mapping is a quadruple---the domain, the range, arcs and edges.
The domain is the abstraction of $A_1\theta$,
the range is the abstraction of $B_1$ and arcs and edges represent
order relations between the nodes of the domain and range. Note that
edges are undirected, while arcs are directed.

We extend this construction by
adding numerical arcs and edges between numerical argument positions. 
These arcs and edges are added if 
numerical inequalities and equalities between the arguments can be deduced.
Deduction of numerical edges and arcs is usually done by considering the 
clauses. However, if a subquery $q$ unifies with a head of a clause
$A\imp B_1,\ldots,B_k,\ldots,B_n$ and we want to know the relation 
between $q$ and $B_k$ (under appropriate substitutions), we {\em may} use the
results of the abstract interpretation to conclude numerical constraints for
$B_1,\ldots,B_{k-1}$. The reason is that if we arrive at $B_k$, this
means that we have proved $B_1,\ldots,B_{k-1}$ 
(under appropriate substitutions). All query-mapping pairs deduced in this 
way are then repeatedly composed. The process terminates because there is
a finite number of query-mapping pairs. 

A query-mapping pair is called {\em circular} if the
query coincides with the range. The initial query terminates if for 
every circular query-mapping pair one of the following conditions
holds:
\begin{itemize}
\item The circular pair meets 
the requirements of the termination test of~\cite{Lindenstrauss:Sagiv}.
\item 
There is a non-negative termination function for which we can prove a decrease
from the domain to the range using the numerical edges and arcs and the 
constraints of the domain and range.
\end{itemize}
Two points should be referred to: how does one automate the guessing of the function
? And how does one prove that it decreases? 
Our heuristic for guessing a termination function is based on the inequalities 
appearing in the abstract constraints.  Each inequality of the form 
{\sl Exp1}$\;\rho\;${\sl Exp2} where $\rho$ is one of $\{>,\geq\}$ suggests 
a function {\sl Exp1-Exp2}. 

The common approach to termination analysis is to find 
{\em one} termination function that decreases over all possible execution paths.
 This leads to complicated termination functions. Our approach allows one
to guess a number of relatively simple termination functions, each suited 
to its query-mapping pair. Since termination functions are simple to find, the 
guessing process can be performed automatically.

After the termination function is guessed, its decrease must be proved.
Let $V_1,\ldots,V_n$ denote numerical argument positions in the domain and
$U_1,\ldots,U_n$ the corresponding numerical argument positions in the range
of the query-mapping pair. First, edges of the query-mapping pair 
are translated to equalities and arcs, to inequalities between these variables.
Second, the atom constraints for the $V$'s
and for the $U$'s are added.
Third, let $\varphi$ be a termination function. We would like to check that
$\varphi(V_1,\ldots,V_n) > \varphi(U_1,\ldots,U_n)$ is implied by the constraints. Thus, we add the negation of this claim to the collection of the 
constraints
and check for unsatisfiability. Since termination functions are linear,
CLP-techniques,
such as CLP(R)~\cite{Jaffar:Maher} and CLP(Q,R)~\cite{CLP:Manual},
are robust enough to obtain the desired contradiction. Note however, 
that if more powerful constraints solvers are used, non-linear termination
functions may be considered.

To be more concrete:
\begin{example}
\label{difficult:loops}
Consider the following program with query {\tt p($i$,$i$)}.
\begin{eqnarray*}
&& \mathtt {p(0,\_).}\\
&& \mathtt {p(X,Y)\imp X>0,X<Y,X1\;\;is\;\;X+1, p(X1,Y).}\\
&& \mathtt {p(X,Y)\imp X>0,X\geq Y,X1\;\;is\;\;X-5, Y1\;\;is\;\;Y-1, p(X1,Y1).}
\end{eqnarray*}

We get, among others, the circular query-mapping pair having the query 
({\tt p($i$,$i$)},$\{\mbox{\sl arg1}>0,\mbox{\sl arg1}<\mbox{\sl arg2}\}$) 
and the mapping given in Figure~\ref{qmdl}. The termination function 
derived for the circular 
query-mapping pair is {\sl arg2}$-${\sl arg1}. In this case, 
we get from the arc and the edge 
the constraints: $V_1<U_1, V_2=U_2$. 
We also have that $V_1>0, U_1>0, V_1<V_2, U_1<U_2$. We would like to prove that
$V_2-V_1>U_2-U_1$ is implied. Thus, we add $V_2-V_1\leq U_2-U_1$ to the set of
constraints and CLP-tools
easily prove unsatisfiability, and thus, that the required implication holds.
\end{example}
\wgefig{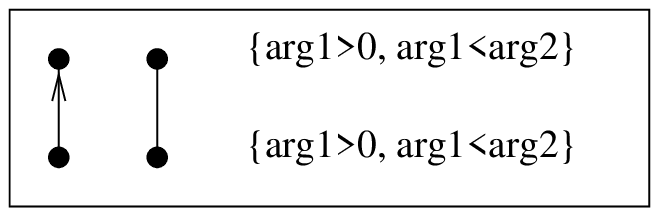}{Mapping for {\tt p}}{qmdl}{0.40\textwidth}

In the case of the 91-function the mappings are given in Figure~\ref{qm91}. 
(We omit the queries from the query-mapping pairs, since
they are identical to the corresponding domains.)
\wgefig{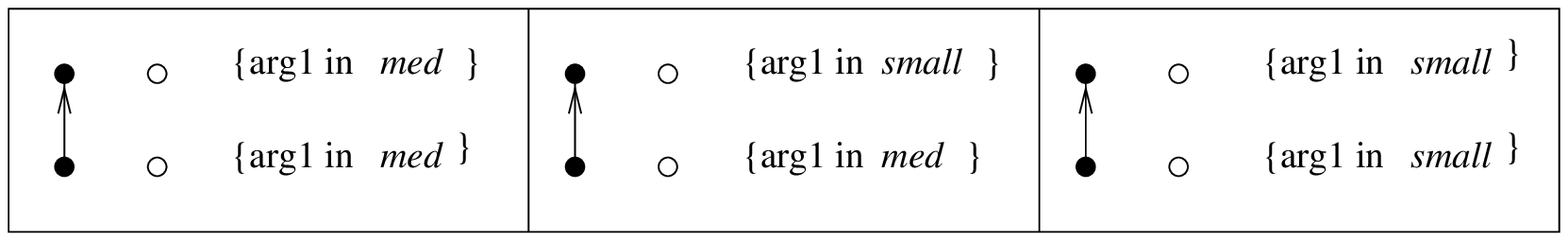}{Mappings for McCarthy's 91 function}{qm91}{0.95\textwidth}

In the examples above there were no interargument relations of the type considered in~\cite{Lindenstrauss:Sagiv}.
However, this need not 
be the case in general. Consider the following program with the query 
{\tt q($b$,$b$,$i$)}. 
\begin{eqnarray*}
&(1)&\mathtt {q(s(X),X,\_).}\\
&(2)&\mathtt {q(s(X),X,N)\imp N>0, N1\;\;is\;\;N-1, q(s(X),X,N1).}\\
&(3)&\mathtt {q(s(s(X)),Z,N)\imp N=<0, N1\;\;is\;\;N-1, q(s(X),Y,N1),q(Y,Z,N1).}
\end{eqnarray*}
Note that constraint inference is an essential step for
proving termination---in order to infer that there is a decrease in the first argument
between the head of (3) and the second recursive call ($s(s(X))\succ Y$ with 
respect to
the norm), one should infer that the second argument in {\tt q} is less than 
the
first with respect to the norm ($s(X)\succ Y$ with respect to the norm). We get
among others circular query-mapping pairs having the mappings presented 
in Figure~\ref{qmq}. The queries of the mappings coincide with the 
corresponding domains. In the first mapping termination follows from the 
decrease in the third
argument and the termination function {\sl arg3}$>0$. 
In the second mapping termination
follows from the norm decrease in the first argument.
\wgefig{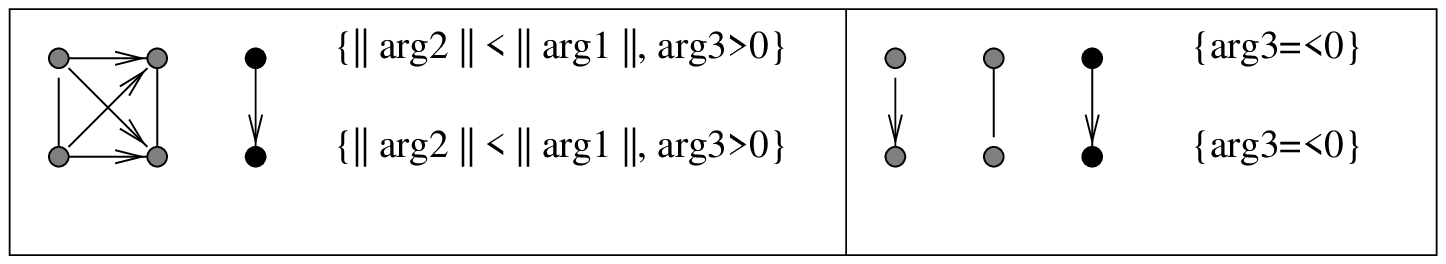}{Mappings for {\tt q}}{qmq}{0.80\textwidth}

\section{The Algorithm}
\label{Algorithm}
In this section we combine all the techniques suggested so far. The complete algorithm {\sf Analyse\_Termination
} is presented in
Figure~\ref{algo}. We use the common notation of {\bf continue} as a keyword,
that can be used only in the body of a looping command and invokes skipping 
to the next iteration of the loop. Each step of the algorithm corresponds to 
one of the previous sections. 

\begin{figure}[htb]
\begin{center}
\fbox{
\parbox{6in}{
\begin{tabbing}
AAAA \= AAAAA\= AAAAA\= AAAAA\= AAAAA\= AAAAA\= AAAAA\=\kill
{\bf Algorithm} \>\> {\sf Analyse\_Termination}\\
{\bf Input}     \>\> A query pattern $q$ and a Prolog program $P$\\
{\bf Output}    \>\> {\sf YES}, if termination is guaranteed\\
                \>\> {\sf NO}, if no termination proof was found\\
\\
\ \,(1) \> {\bf If} {\sf TermiLog\_Algorithm($P$,$q$)=YES} {\bf then}\\
\ \,(2) \>      \>{\bf Return} {\sf YES}.\\
\ \,(3) \> {\bf If} there is no numerical loop in $P$ {\bf then}\\
\ \,(4) \>      \>{\bf Return} {\sf NO}.\\
\ \,(5) \> {\bf Guess} and {\bf verify} numerical argument positions;\\
\ \,(6) \> {\bf Compute} integer abstraction domain;\\
\ \,(7) \> {\bf Compute} abstractions of answers to queries (optional);\\
\ \,(8) \> {\bf Compute} query-mapping pairs;\\
\ \,(9) \> {\bf For each} circular query-mapping pair {\bf do}:\\
(10) \>         \> {\bf If} its circular variant has a forward positive cycle {\bf then}\\
(11) \>      \>      \> {\bf Continue};\\
(12) \>         \>{\bf Guess} bounded (integer-valued) termination function;\\
(13) \>         \>{\bf Traverse} the query-mapping pair {\bf and  compute} 
values\\
     \>         \>      \>of the termination function;\\
(14) \>         \>{\bf If} the termination function decreases monotonically {\bf then}\\ 
(15) \>         \>      \> {\bf Continue};\\
(16) \>         \> {\bf Return} {\sf NO}; \\
(17) \> {\bf Return} {\sf YES}.\\
\end{tabbing}
}}
\end{center}
\caption{Termination Analysis Algorithm}
\label{algo}
\end{figure}

Note that Step 7, computing the abstractions of answers to queries, is 
optional. If the algorithm returns {\sf NO}
it may be re-run either with Step 7 included or with a different integer 
abstraction domain. 

The {\sf Analyse\_Termination} algorithm is sound:
\begin{theorem}
Let $P$ be a program and $q$ a query pattern.
\begin{itemize}
\item {\sf Analyse\_Termination($P$, $q$)} terminates.
\item If {\sf Analyse\_Termination($P$, $q$)} reports {\sf YES} then,
for every query $Q$ matching the pattern $q$, the LD-tree of $Q$ w.r.t.\ $P$ is finite.
\end{itemize}
\end{theorem}

\section{Conclusion}
\label{Conclusion}
We have shown an approach that can extend the scope of existing automatic 
systems for proving termination to programs containing arithmetic predicates.
To do so we first introduced a new kind of constraints. Second, we indicated how
such constraints may be inferred. Finally, we showed that the query-mapping
pairs technique is robust enough to incorporate them by using very simple
termination functions that are derived from the constraints.

We have not yet implemented the ideas discussed in this paper, but it is
clear that even a simple implementation, without inference of comparisons,
unfolding and application of abstract interpretation as suggested in 
Section~\ref{AbstrInterp}, will be very useful from a practical point of view,
since many
programs encountered in practice cannot be handled by systems that use norms to
prove termination because of very simple numerical loops.

\bibliography{
/home/alexande/M.Sc.Thesis/main}
\bibliographystyle{abbrv}
\end{document}